# Polaron transport and thermoelectric behavior in La-doped SrTiO$_3$ thin films with elemental vacancies

*Woo Seok Choi\*, Hyang Keun Yoo, and Hiromichi Ohta*

Prof. Woo Seok Choi
Department of Physics, Sungkyunkwan University, Suwon 440-746, Korea
E-mail: choiws@skku.edu
Dr. Hyang Keun Yoo
Department of Physics and Astronomy, Seoul National University, Seoul 151-747, Korea
Center for Correlated Electron Systems, Institute for Basic Science, Seoul 151-747, Korea
Prof. Hiromichi Ohta
Research Institute for Electronic Science, Hokkaido University, Sapporo 001-0020, Japan

Electrodynamic properties of La-doped SrTiO$_3$ thin films with controlled elemental vacancies have been investigated using optical spectroscopy and thermopower measurement. In particular, we observed a correlation between the polaron formation and thermoelectric properties of the transition metal oxide (TMO) thin films. With decreasing oxygen partial pressure during the film growth ($P(O_2)$), a systematic lattice expansion was observed along with the increased elemental vacancy and carrier density, experimentally determined using optical spectroscopy. Moreover, we observed an absorption in the mid-infrared photon energy range, which is attributed to the polaron formation in the doped SrTiO$_3$ system. Thermopower of the La-doped SrTiO$_3$ thin films could be largely modulated from -120 to -260 $\mu$V K$^{-1}$, reflecting an enhanced polaronic mass of ~3 < $m_{polron}/m$ < ~4. The elemental vacancies generated in the TMO films grown at various $P(O_2)$ influences the global polaronic transport, which governs the charge transport behavior, including the thermoelectric properties.

# 1. Introduction

Transition metal oxides (TMOs) have been drawing lots of recent attention as a promising candidate for highly efficient thermoelectric material with novel mechanisms.[1,2] Owing to the strongly correlated nature of the *d*-electrons, the TMOs exhibit intriguing physical properties including enhanced thermopower and high electrical conductivity. For example, a large thermopower has been observed in $Na_xCoO_2$ due to the strong correlation.[3] In ruthenate superlattices, spin channel dependent thermopower and large thermopower anisotropy have been theoretically predicted.[4,5] Many studies have also focused on the 3*d* perovskite $SrTiO_3$ (STO) based material system, which is an *n*-type thermoelectric material.[6-19] A dimensional crossover of thermoelectric phenomena has been observed in Nb-doped $SrTiO_3$ based heterostructure,[6] where the $3d^1$ electrons in Ti strongly couple with the lattice phonons to exhibit a large thermopower.[7,8] Also, in similar titanate oxide heterostructures ($LaTiO_3/SrTiO_3$ superlattices), fine tuning of the thermopower has been achieved using fractional doping of the 2D carriers.[9]

Generally, thermopower of TMO is closely related to the *d*-orbital electronic structure near the Fermi level ($E_F$), as all the other electronic transport properties.[4,10] Particularly, electronic structure modification due to the polaron formation can significantly affect the thermoelectric properties of TMOs,[7,8,20] although not much attention has been paid to the polaron transport for most of the thermoelectric materials. A polaron is a quasi-particle which can be described as a coupling of electron and phonon. The strong coupling usually dresses the otherwise free electron, and induces changes in the electrodynamic and optical properties. In particular, for an electron under a strong electron-phonon coupling, an enhanced inert mass, $m_{polaron}$ can be expected due to the impeded motion of the electron. Since enhanced effective mass usually increases the thermopower, we can expect larger thermopower for systems exhibiting polaron transport.

Here we present electrodynamic properties of 3$d$ TMO thin films, by modifying the elemental vacancy concentration. In particular, using optical spectroscopy, we investigated polaron transport of La-doped SrTiO$_3$ (LSTO) thin films with elemental vacancies induced by a systematic growth of the films at various oxygen partial pressures ($P(O_2)$). The modification in the elemental vacancy concentration resulted in a systematic change in the crystal structure and electrodynamic properties. The polaron effective mass enhancement and the thermopower showed a maximum for an intermediate $P(O_2)$ values, implying that the elemental vacancy engineering using thin film growth is an efficient method for controlling the thermopower of TMO thin films.

## 2. Experiments

La$_{0.15}$Sr$_{0.85}$TiO$_3$ (LSTO) thin films were heteroepitaxially grown on (001)-oriented single-crystalline (LaAlO$_3$)$_{0.3}$(SrAl$_{0.5}$Ta$_{0.5}$O$_3$)$_{0.7}$ (LSAT) substrates by pulsed laser deposition (PLD) technique. We used a KrF excimer laser ($\lambda = 248$ nm) with a repetition rate of 1 Hz to ablate the stoichiometric LSTO ceramic target. We used 15% of La doping to induce high carrier concentration, for viable observation using optical spectroscopy. The laser energy density and the deposition temperature were 1.75 J cm$^{-2}$ and 500°C, respectively. We controlled $P(O_2)$ from $1 \times 10^{-2}$ to $1 \times 10^{-6}$ Torr, which results in a systematic change in the crystal lattice and electronic structures. **Figure 1(a)** shows the x-ray diffraction (XRD, CuK$\alpha_1$) $\theta$-$2\theta$ scan of the LSTO films on LSAT prepared at various $P(O_2)$. Intense Bragg diffraction peaks of the 002 LSTO are observed together with the Bragg diffraction peak of the 002 LSAT substrates. All the films show well-defined Kiessig and Pendellösung fringes, indicating homogeneous flatness and good out-of-plane orientation of the films. The thicknesses of the films were around 180 nm, calculated using the fringes. With decreasing $P(O_2)$, a systematic increase in the out-of-plane lattice constant is observed as shown in Figure 1(b).

The in-plane reflectance spectra as a function of photon energy ($R(\omega)$) of the LSTO thin films on LSAT substrate in the near-normal incident geometry were measured between 3.7 meV and 6.9 eV using Fourier transform infrared spectrophotometers and a grating monochromator. An LSAT substrate was chosen for viable optical spectroscopic investigation with minimized lattice mismatch. While STO substrates should have the smallest lattice mismatch with the LSTO thin films, the chemical contrast between the two is rather weak. The similarity in the crystal and electronic structures makes it very difficult to distinguish between the thin film and the substrate optically, in the photon wavelength range of interest, where the penetration depth is in the order of a few tens or hundreds of nanometers.

## 3. Results and Discussion

**Figure 2** shows $R(\omega)$ of the LSTO thin films measured at room temperature. $R(\omega)$ of the bare LSAT substrate has been included for comparison. Note that the optical property of LSAT substrate could change during the film deposition process. We indeed observed a modest change at the high photon energy (> 3 eV) in the transmittance spectra of the substrate by annealing it at the growth temperature and $P(O_2)$. However, we confirmed that $R(\omega)$ showed negligible changes (within 1%), suggesting that $R(\omega)$ of bare LSAT substrate can be used as the reference. Due to the large penetration depth of the probing light, as mentioned above, $R(\omega)$ reflect both the responses of the films and substrates.

The spectral features of $R(\omega)$ below 0.1 eV are mainly due to the infrared active phonons of the LSAT substrate. The responses from the first two phonons of the perovskite structure of LSTO film located at ~0.01 and ~0.02 eV could not be well distinguished due to overlapping with LSAT phonon responses. However, the response from the third phonon located at ~0.07 eV could be clearly distinguished from $R(\omega)$ of the substrate. Starting from $R(\omega)$ of LSAT, the difference between the peak and valley of the spectra decreases, and all the spectral features below ~0.1 eV shifts slightly to the lower energy, systematically with decreasing $P(O_2)$. This is due to the increased number of free carriers generated in the LSTO films with decreasing $P(O_2)$, which effectively screen the response from the optical phonons.

The spectral features of $R(\omega)$ above 0.1 eV can be attributed to three different contributions. Above ~3.2 eV, which is the bandgap of STO (and LSTO), the charge-transfer transition from O $2p$ to Ti $3d$ dominates the spectral characteristics. Between ~1 and ~3.2 eV, the film is largely transparent, and the Fabry-Perot oscillation determined by Fresnel equation is responsible for the features. Finally, between ~0.1 and ~1 eV, there are additional optical absorptions in the films, due to the electron-phonon coupling [7,21-23] and generation of the elemental vacancies.[24]

In order to obtain quantitative information on the electrodynamic properties and electronic band structures of the LSTO films, we calculated $R(\omega)$ by assuming that the in-plane optical conductivity spectra ($\sigma_1(\omega)$) of the film could be represented by a Drude and several Lorentz oscillators:

$$\sigma_1(\omega) = \frac{e^2}{m^*} \frac{n_D \gamma_D}{\omega^2 + \gamma_D^2} + \frac{e^2}{m^*} \sum_j \frac{n_j \gamma_j \omega^2}{\left(\omega_j^2 - \omega^2\right)^2 + \gamma_j^2 \omega^2} \qquad (1).$$

Here, $e$ and $m^*$ are the electronic charge and the effective mass, respectively. In addition, $n_j$, $\gamma_j$ and $\omega_j$ are the carrier density, the scattering rate, and the resonant frequency of the $j$-th oscillator, respectively. The first term describes the coherent Drude response from free charge carriers denoted with subscript $D$, and the second term describes the incoherent response due to bound electronic states.[25]

The empty symbols (circles) in Figure 2(b) present the experimental spectra of LSTO film grown at $P(O_2) = 1 \times 10^{-2}$ Torr, to illustrate the Drude-Lorentz curve-fitting of $R(\omega)$. The simulated $R(\omega)$ based on different Drude-Lorentz oscillators are shown in lines: The dotted line represents simulated $R(\omega)$ using one Drude oscillator, and two Lorentz oscillators, both centered above 3.5 eV, which defines the charge-transfer gap of STO. Within this simple model we note significant discrepancy between the simulation and experimental result in the range between ~0.2 and ~0.6 eV. To compensate the discrepancy, one more Lorentz oscillator located at ~0.4 eV is necessary (solid line), which results in excellent agreement with the experimental $R(\omega)$. In addition, for samples grown at lower $P(O_2)$, another Lorentz oscillator located at ~1.4 eV was necessary to properly fit the experimental $R(\omega)$.

The result of Drude-Lorentz fitting of $R(\omega)$ is presented in **Figure 3** as optical conductivity spectra ($\sigma_1(\omega)$) of LSTO thin films. The spectral features above the charge-transfer gap are almost the same for all the films, suggesting the electronic structures regarding the bandgap do not change much with $P(O_2)$ (Figure 3(a)). However, the low energy features shows significant and systematic changes with $P(O_2)$. As shown in Figure 3(b), the low photon energy $\sigma_1(\omega)$ ($\omega < $ ~2.5 eV) could be characterized by one Drude and two Lorentz oscillators. As an example, $\sigma_1(\omega)$ of the LSTO thin film grown at $1 \times 10^{-4}$ Torr has been deconvoluted into the individual oscillators. The peak centered at the origin (dark grey) is the Drude oscillator and shows systematic increase with decreasing $P(O_2)$. The first Lorentz oscillator is located at ~0.4 eV (grey), in the mid-infrared (MIR) photon energy range, and is responsible for the feature in $R(\omega)$ (Figure 2(a)) between ~0.1 and ~1 eV. As mentioned previously, we believe that this MIR peak is coming from the polaron formation in the STO-based film. The second Lorentz oscillator located at ~1.4 eV (light grey) shows a systematic increase with decreasing $P(O_2)$. This peak is known to come from the vacancy absorptions (VA), which is related to both cation (Sr) and oxygen vacancies.[24,26,27]

The formation of polaron can be optically observed by the renormalization of the coherent Drude absorption in $\sigma_1(\omega)$, where some of its spectral weight ($SW \equiv \int \sigma_1(\omega) \, d\omega$) shifts to higher energies (typically in the MIR photon energy region) as multi-phonon absorption bands.[28-30] The MIR peak structure, and hence the polaron formation has frequently been observed in different doped STO systems,[7,21,22,31] and it is apparent that the transport properties of our LSTO thin films are influenced by the polaron formation as well.

Based on the deconvoluted absorption peaks shown in Figure 3, we performed $SW$ analyses as a function of $P(O_2)$ (**Figure 4(a)**). As expected, the $SW$ of the Drude absorption ($SW_{Drude}$) decreases with increasing $P(O_2)$, indicating that the film becomes more metallic when grown at low $P(O_2)$. In particular, the carrier densities ($n_D$) of the LSTO thin films, which is related to the plasma frequency ($\omega_p^2 = ne^2/\varepsilon_0 m^*$), is shown in Figure 4(b). The values have been obtained from the Drude-Lorentz fitting, using $m^*/m_e = 6$.[15,29] $n_D$ values of our thin films range between ~4 × $10^{20}$ and ~1 × $10^{21}$ cm$^{-3}$, which is similar to other LSTO samples.[11,15,17,18] The decreasing trend of $n_D$ with increasing $P(O_2)$ was consistently observed. Similarly, the $SW$ of the vacancy absorption ($SW_{VA}$) also showed a monotonic decrease with increasing $P(O_2)$. Such a trend in the VA peak has also been observed in insulating $SrTiO_{3-\delta}$ thin films.[24] While our LSTO films are metallic for all $P(O_2)$ studied due to the La doping, the increase of $SW_{VA}$ suggests that both Sr and oxygen vacancy increases systematically with decreasing $P(O_2)$, which will influence the charge transport properties of the films. Finally, the *SW* of the MIR peak ($SW_{MIR}$) does not show a monotonic trend: With increasing $P(O_2)$, $SW_{MIR}$ increases, but it shows a maximum at $P(O_2) = 10^{-4}$ Torr, and then decreases. The behavior of $SW_{MIR}$ as a function of $P(O_2)$ suggests that the polaron behavior cannot be simply explained by the conductivity (e.g., carrier density or electron mobility), elemental vacancies, or out-of-plane lattice elongation (Figure 1).

The polaron effective mass $m_{polaron}$ can be quantitatively determined using the *SW* analyses ($m_{polaron}/m = (SW_{Drude} + SW_{MIR}) / SW_{Drude}$).[7,22] For slightly doped bulk $SrTi_{1-x}Nb_xO_3$ ($0 \leq x \leq 0.02$), the *SW* analyses resulted in $m_{polaron}/m = 2.0 \pm 0.3$, regardless of the doping rate.[22] On the other hand, a dimensional crossover from 3D to 2D was responsible for a significant enhancement of $m_{polaron}/m$, which increased up to $5.9 \pm 0.4$ for the single layer of Nb-doped STO.[7] For our LSTO thin films, $m_{polaron}/m$ exhibited slightly larger values than the bulk, as shown in **Figure 5(a)**. As predicted from $SW_{MIR}$ behavior, $m_{polaron}/m$ shows a maximum for $P(O_2) = 10^{-4}$ Torr. From analyses of the Drude peak in $\sigma_1(\omega)$, we can also obtain the relaxation time ($\tau$), which is the inverse of the scattering rate (half-width-half-maximum of the peak, $\gamma$): $\tau = 1/\gamma$. Figure 5(b) shows the relaxation time of the carriers in LSTO thin films as a function of $P(O_2)$. At low $P(O_2)$ regime $\tau$ increases with increasing $P(O_2)$. However, above $P(O_2) = 3 \times 10^4$ Torr, $\tau$ decreases again with increasing $P(O_2)$.

Figure 5(c) shows the thermopower (*S*) values of the LSTO thin films at room temperature, as a function of $P(O_2)$, measured by a conventional steady state method, which introduced a temperature gradient in the in-plane direction. A thermo-electromotive force ($\Delta V$) is generated between the two ends of the LSTO films by introducing a temperature difference ($\Delta T$). The overall *S* values of our LSTO films are comparable to other doped STO bulk, thin films, and heterostructures.[6,7,9,12-18] However, the LSTO samples grown at different $P(O_2)$ did not show the usual increasing trend of $S$ with decreasing $n_D$, which has been observed in most of the previous studies.[7,9,11-14] On the contrary, $S$ of our LSTO samples show a maximum value for $P(O_2) = 10^{-5}$ Torr (~260 $\mu$V/K).

It is notable that $m_{polaron}/m$, $\tau$, and $S$ all show similar $P(O_2)$-dependent trend, with a maximum at an intermediate $P(O_2)$ value. In particular, instead of monotonically decreasing with increasing $n_D$ (or $P(O_2)$), $S$ shows a peak behavior which is consistently shown for $m_{polaron}/m$ and $\tau$. This indicates that the polaron transport in the LSTO film governs the overall electronic transport behavior including the thermopower. In particular, $m_{polaron}/m$, $\tau$, and $S$ should show linear dependencies with each other within the polaron transport scheme.[7,30,32] Hence, our observation suggests that strong electron-phonon coupling is indeed beneficial for the large thermopower and thermoelectric efficiency and the polaron transport should be considered in understanding the electrodynamic and thermoelectric properties of LSTO thin films.

We further note that $m_{polaron}/m$, $\tau$, and $S$ show maximum at a slightly different $P(O_2)$ value of $10^{-4}$, $3 \times 10^{-5}$, and $10^{-5}$ Torr, respectively, while the overall $P(O_2)$-dependent behavior is similar. This implies that the detailed transport properties can be tuned by elemental vacancies, within the polaron transport behavior. In particular, both the expansion of the out-of-plane lattice constant (Figure 1) and the enhancement of $SW_{VA}$ (Figure 4(a)) coherently indicate the increase of the cation (La and Sr) and oxygen vacancies with decreasing $P(O_2)$. The increase of the oxygen vacancies is evident as less oxygen should be incorporated into the film during the growth in the films grown at low $P(O_2)$. On the other hand, it has been suggested that the increase in the lattice constant of the STO-based films is largely due to the cation (Sr) vacancies, which is created along with the oxygen vacancies.[33-39] While it is not clear how the La and Sr vacancies are distributed within our LSTO films, it is expected that the cation vacancies also increase with decreasing $P(O_2)$.

The two different types of elemental vacancies play distinctive roles in determining the transport properties of the LSTO films. Oxygen vacancies generally add two electrons in the Ti site, providing charge carriers to the normally insulating STO. The increasing $SW_{Drude}$ and $n_D$ with decreasing $P(O_2)$ in Figure 4 supports this argument. The increase of $n_D$ for the LSTO films grown at $P(O_2) \geq 1 \times 10^{-4}$ Torr accompany increase in the $m_{polaron}/m$, $\tau$, and $S$. On the other hand, LSTO films grown at $P(O_2) \leq 3 \times 10^{-5}$ Torr experience significantly larger defect scattering due to the enhanced cation vacancies. The cation vacancies are known to compensate the generation of carriers by the oxygen vacancies. In addition, it also expands the lattice, thereby disturbing the carrier transport (Figure 1). The sharp drop in $\tau$ for low $P(O_2)$ clearly suggests the defect scattering with increased cation vacancies, which in turn decreases $m_{polaron}/m$ and $S$ within the polaron transport scheme.

We finally note that the polaron mass enhancement is different from the electronic band mass, conventionally determined by the curvature of the electronic band.[22] Therefore, our discussion on the polaronic conduction cannot be directly applied to the argument regarding the $P(O_2)$ dependent change in the effective mass in doped STO.[17,40] However, as we have shown in this paper, the polaron transport in LSTO films provide critical understanding on the system, which should be carefully considered in discussing the thermoelectric properties.[20]

## 4. Conclusion

We have studied the electrodynamic properties of epitaxial LSTO thin films grown at different oxygen partial pressures. The LSTO thin films obey the polaron transport behavior with an enhanced polaronic mass of ~3 < $m_{polaron}/m$ < ~4. The vacancy and charge carriers generated at low oxygen partial pressure growth determined the detailed transport properties within the polaron transport scheme, suggesting that the elemental vacancy engineering in transition metal oxide thin films is a viable approach for efficient thermoelectric materials.

**Acknowledgements**
This work was supported by Basic Science Research Program through the National Research Foundation of Korea (NRF) funded by the Ministry of Science, ICT and future Planning (NRF-2014R1A2A2A01006478). HYK is supported by the Institute for Basic Science in Korea. HO is supported by JSPS-KAKENHI (25246023, 25106007).

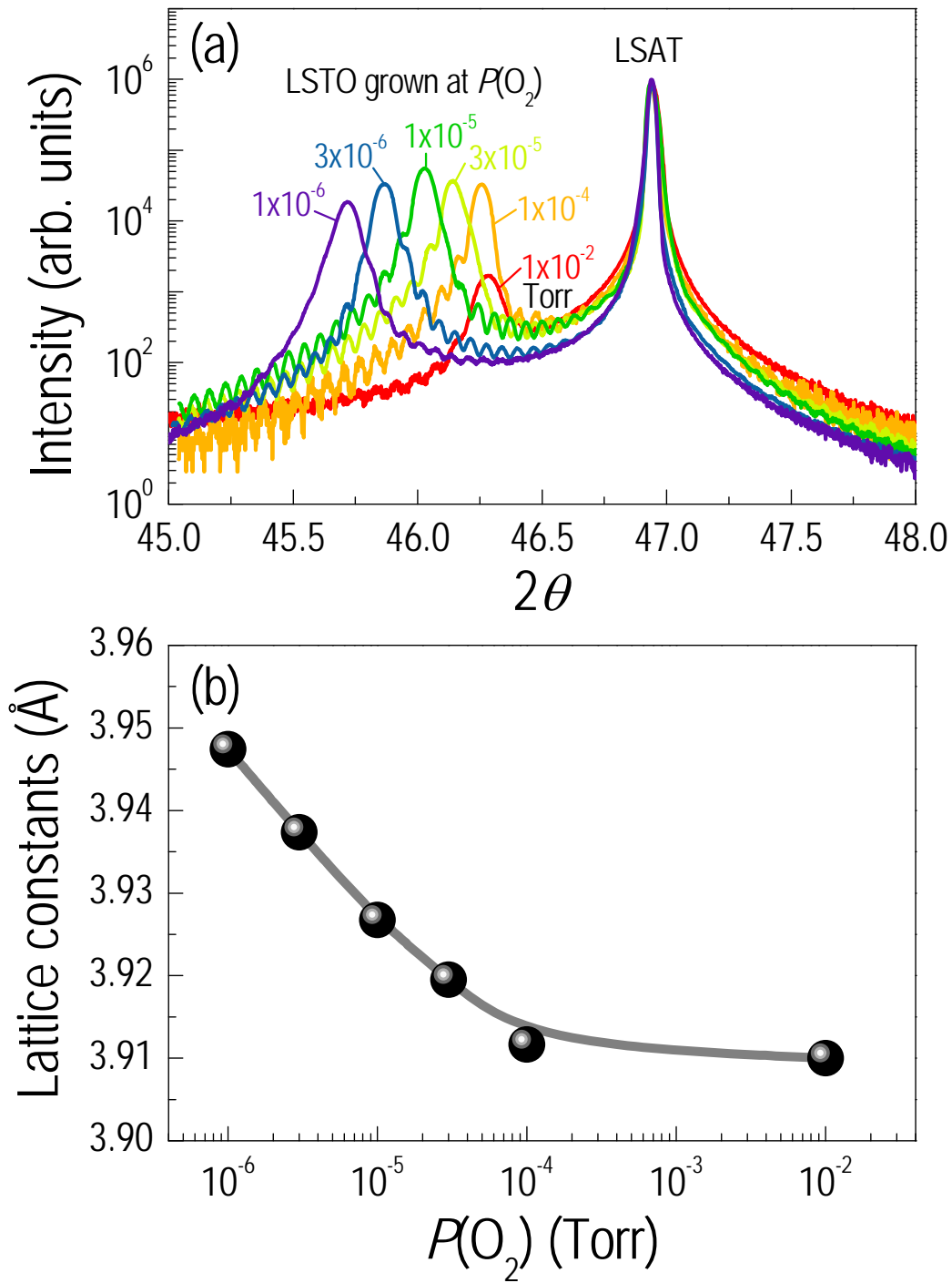

**Figure 1.**
X-ray diffraction patterns of LSTO thin films on LSAT substrates. a) Out-of-plane XRD $\theta$-$2\theta$ scan around 002 Bragg diffraction of the LSTO thin films grown on LSAT at different $P(O_2)$. As the $P(O_2)$ decreases, the LSTO peak shifts to the lower $2\theta$ angle. b) Systematic increase of the out-of-plane lattice constant with decreasing $P(O_2)$. The error bars are smaller than the symbols. The grey line is drawn to guide the eye.

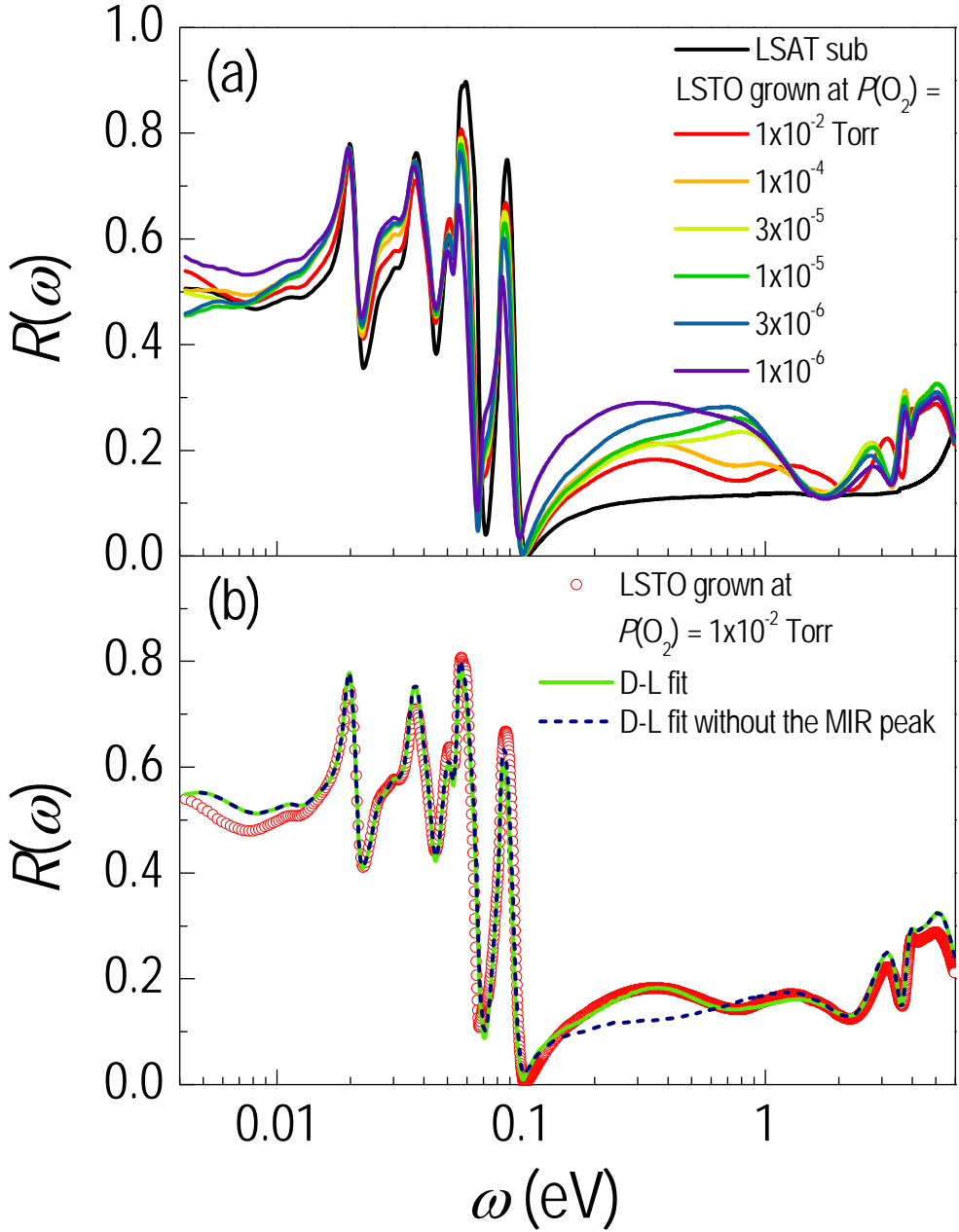

**Figure 2.**
$R(\omega)$ of the LSTO thin films and model fit result. a) Room temperature $R(\omega)$ of LSAT substrate and LSTO thin films on LSAT substrate grown at different $P(O_2)$. The systematic changes in the spectral features of the LSTO thin films at low photon energy region (< 0.1 eV) is mostly due to the free carrier contribution in the films. Several unexpected modifications also emerge at higher photon energy region for the LSTO films, compared to $R(\omega)$ of the LSAT substrate. b) Experimental $R(\omega)$ of LSTO thin film grown at $P(O_2) = 10^{-2}$ Torr (symbols), and fitting of $R(\omega)$ using Drude-Lorentz fitting (lines). The solid (dotted) line represents the Drude-Lorentz fitting with (without) a Lorentz oscillator in the MIR photon energy range near 0.4 eV. The MIR absorption is absolutely necessary to fit the experimental data properly.

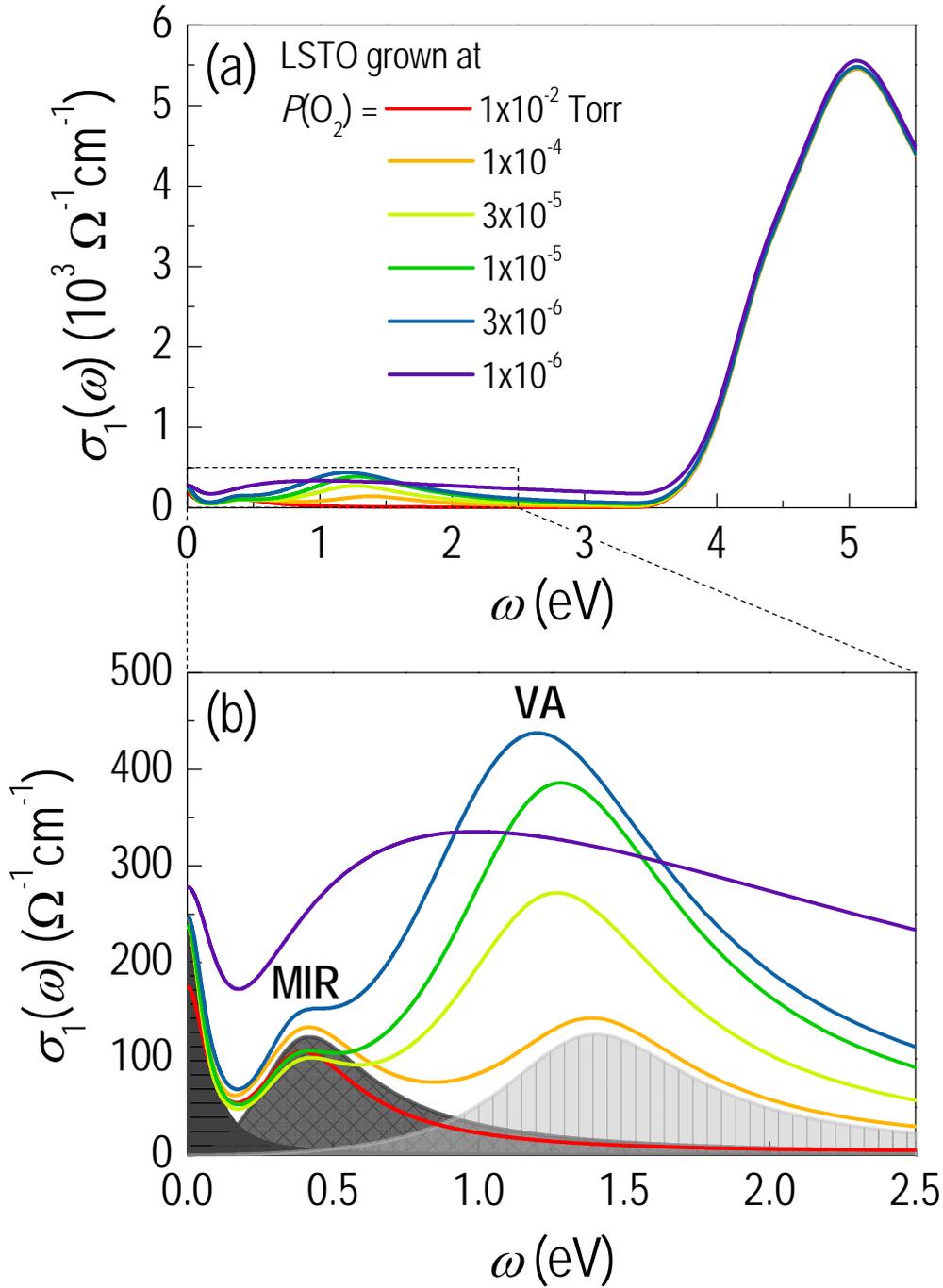

**Figure 3.**
$\sigma_1(\omega)$ and electronic structure of LSTO thin films. a) $\sigma_1(\omega)$ (results from the Drude-Lorentz fit of $R(\omega)$) of LSTO thin films grown on LSAT at different $P(O_2)$. While the high photon energy (> 3.5 eV) features are rather similar for all the samples, the low photon energy feature changes drastically with decreasing $P(O_2)$ at growth. b) Magnified $\sigma_1(\omega)$ below 2.5 eV. $\sigma_1(\omega)$ of the film grown at $P(O_2) = 1 \times 10^{-4}$ Torr is deconvoluted into individual oscillators. Three absorption features (oscillators), *i.e.* Drude (dark grey), MIR (grey), and VA (light grey) peaks centered at the origin, ~0.4 eV, and ~1.4 eV, respectively, could successfully account for the quantitative optical properties of LSTO thin films at low photon energy region. The shaded regions of each peak correspond to the spectral weight of each oscillator.

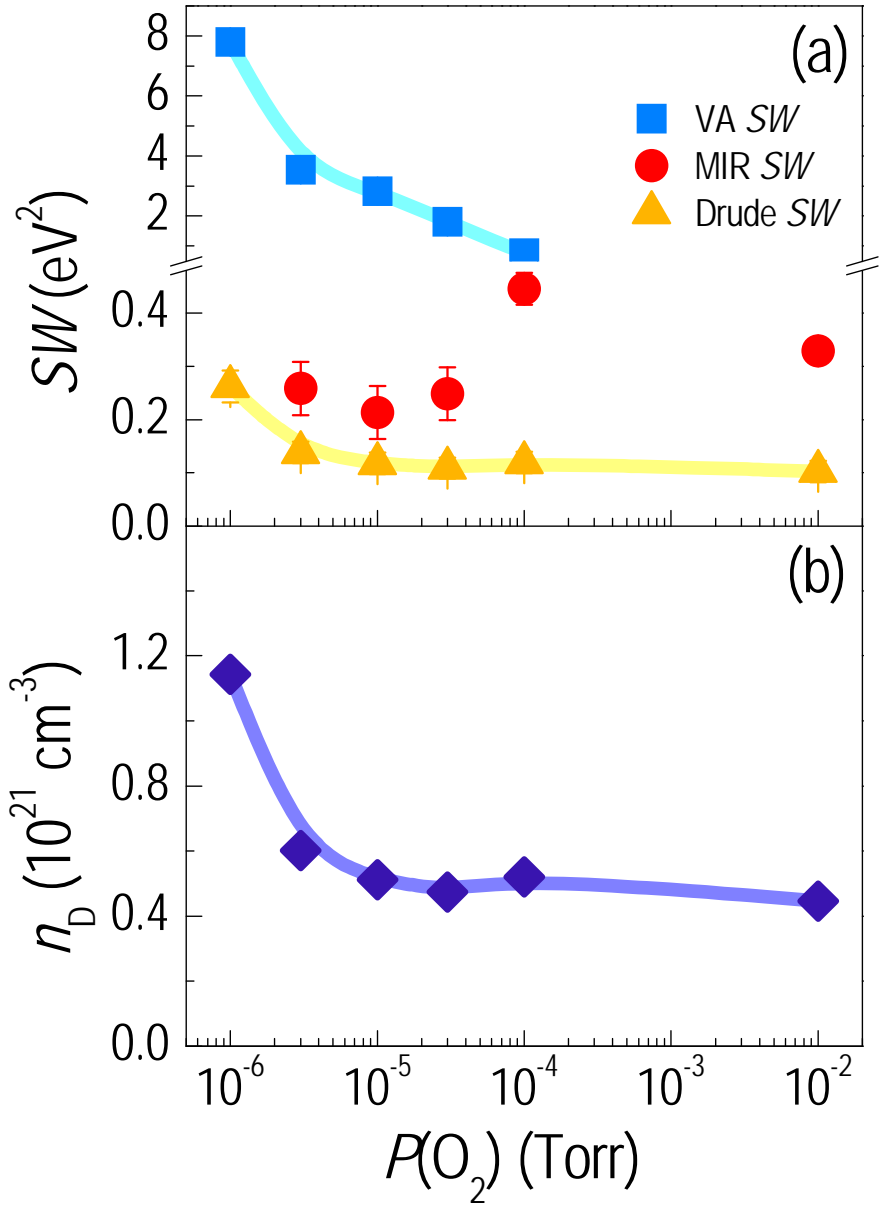

**Figure 4.**
Spectral weight analyses and transport properties of LSTO thin films. a) Resultant spectral weights (*SW*) of VA (squares), MIR (circles), and Drude (triangles) absorption peaks from the Drude-Lorentz fitting as a function of $P(O_2)$. b) Carrier density, *n*, obtained from the Drude analyses as a function of $P(O_2)$. $SW_D$, $SW_{VA}$, and *n* show the same increasing trend with decreasing $P(O_2)$. For the symbols without any error bars, the error bars are smaller than the symbols. The lines are drawn to guide the eye.

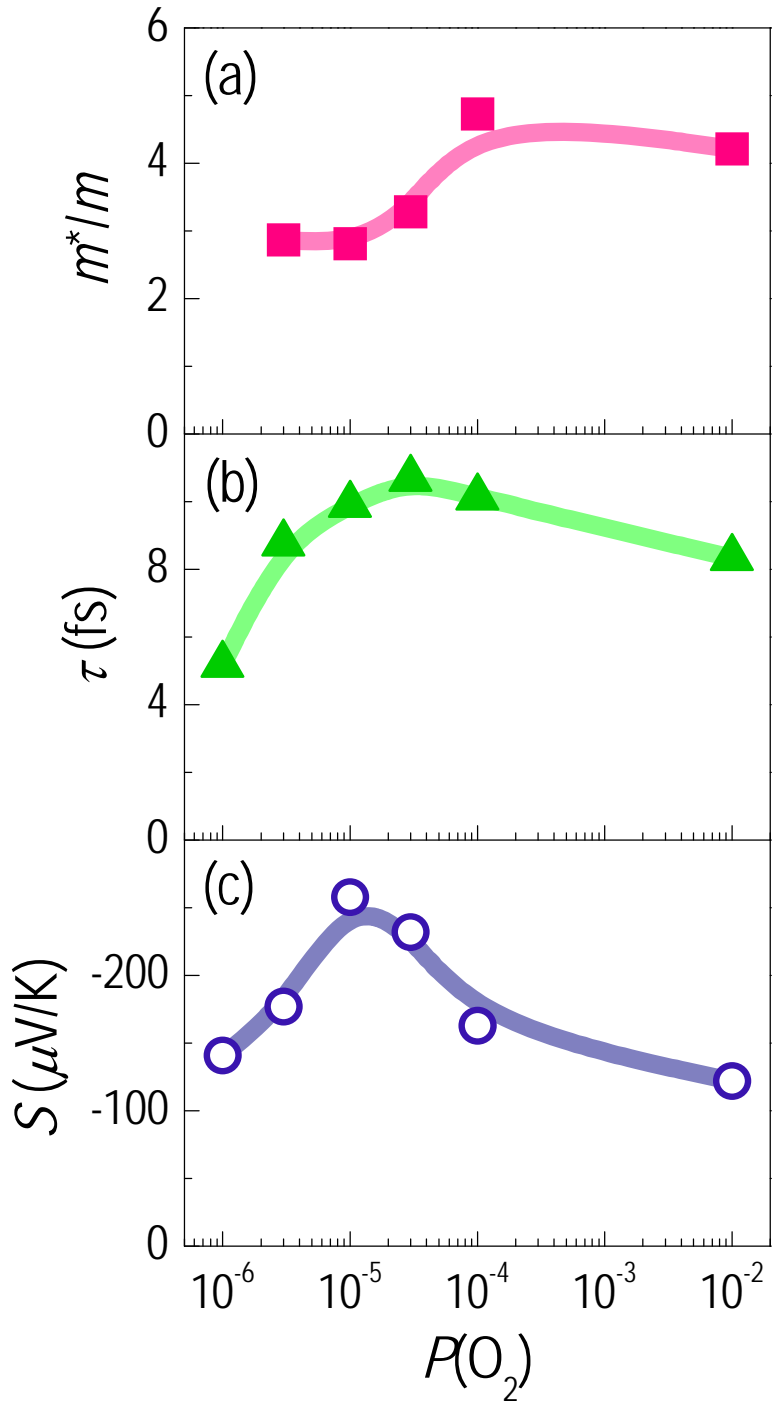

**Figure 5.**
Summary of electrodynamic properties of LSTO thin films grown at various $P(O_2)$. a) $m_{polaron}/m$, b) $\tau$, and (c) $S$ of LSTO films grown at various $P(O_2)$. $m_{polaron}/m$ has been calculated from $m_{polaron}/m = (SW_D + SW_{MIR}) / SW_D$. These three electrodynamic parameters all show different maximum $P(O_2)$. This suggests that various aspects including polaron behavior, contribution from the vacancy energy levels and defect scattering, and crystallographic structure should be considered in understanding the transport mechanism of the LSTO films. The error bars are smaller than the symbols. The lines are drawn to guide the eye.